\documentstyle[aps,prl,twocolumn,epsfig]{revtex}
\newcommand{\beq}{\begin{equation}}
\newcommand{\eeq}{\end{equation}}
\newcommand{\bea}{\begin{eqnarray}}
\newcommand{\eea}{\end{eqnarray}}
\newcommand{\beas}{\begin{eqnarray*}}
\newcommand{\eeas}{\end{eqnarray*}}
\newcommand{\ket}[1]{| #1 \rangle}
\newcommand{\bra}[1]{\langle #1 |}

\newcommand{\pdag}{{\phantom{\dagger}}}

\begin{document}
\draft
\wideabs{
\title{Generalized Numerical Renormalization Group 
for Dynamical Quantities}
\author{Walter Hofstetter}
\address{Theoretische Physik III, Elektronische Korrelationen und
Magnetismus, Universit\"at Augsburg, 86135 Augsburg, Germany}
\date{\today}
\maketitle
\begin{abstract}
In this paper we introduce a new approach for calculating dynamical
properties within the \mbox{numerical} renormalization group. 
It is demonstrated that the method previously used fails for the 
Anderson impurity in a magnetic field due to the absence of 
energy scale separation. 
The problem is solved by evaluating the Green function with respect 
to the reduced density matrix of the \emph{full} system, leading 
to accurate spectra in agreement with the static magnetization. 
The new procedure (denoted as \mbox{DM-NRG}) provides a unifying framework for calculating 
dynamics at any temperature and represents the correct extension 
of Wilson's original thermodynamic calculation.
\end{abstract}
}
Quantum impurity models and their low-temperature properties are 
of central importance in condensed matter physics. 
They show characteristic many-body effects like the screening 
of a local moment by conduction electrons (the Kondo effect) 
which was first observed in measurements on dilute magnetic impurities 
in metals (see \cite{Hewson 93}).
More recently, artificial nanostructures (quantum dots \cite{Goldhaber 
98} or surface atoms probed by STM \cite{Li 98,Manoharan 00}) 
with tunable parameters provided new representations
of the Anderson or Kondo model \cite{Anderson 61,Kondo 64}. 
In theory, a very fruitful line of research was opened by the development 
of dynamical mean-field theory (DMFT) \cite{Metzner 89} where correlated lattice systems 
are mapped onto effective impurity models which are then accessible 
in a controlled way \cite{Georges 96}. 

In all the above areas, progress depends sensitively on the existence 
of a reliable calculational method that can provide static and 
dynamic (spectral) properties in the full energy range.
Wilson's numerical renormalization group \cite{Wilson 75} gave the first 
quantitative description of the Kondo effect. In systems with 
very different energy scales (small Kondo temperature $T_K$,  
large bandwidth) it is the only technique that can do so. 
In the original calculation Wilson focused on obtaining thermodynamic 
expectation values like the impurity susceptibility by iterative 
diagonalization. Each iteration step was shown to correspond to 
a certain temperature where expectation values could be obtained 
with great precision. Later, the method was extended to zero temperature 
dynamical properties by several groups and applied to a variety of 
problems \cite{Sakai 89,Costi 94}, including recent DMFT calculations 
(e.g.~\cite{Bulla 98,Hofstetter 00}). 
In these calculations the additional assumption had to be made that transitions 
from the \emph{ground state} to higher excitations are already correctly 
described in the first few iterations. 
Accurate results in agreement with sum rules 
were obtained for the single particle spectrum in the 
absence of external fields.  
In the following, however, we demonstrate that 
this procedure is not rigorous and fails for 
the Anderson impurity model in a magnetic field. 
To remedy the defect, we introduce a new approach  
based on the concept of the \emph{reduced density matrix}.  
This procedure (which in the following will be denoted as 
DM-NRG) makes use of the full information 
contained in iterative diagonalization and can therefore be considered 
as the true extension of Wilson's original work to dynamical 
quantities. 

To be specific, we consider the spin $1/2$ Anderson model 
$H = H_0 + H_{\rm imp}$ where the impurity part is given by
\bea  \label{H_impurity}
H_{\rm imp} = V\left(f^\dagger_{\sigma} c^\pdag_{0 \sigma} + h.c. \right) 
+ U n_{f \uparrow} n_{f \downarrow} -\epsilon_f n_{f}  
&& - h\,S_f^z.
\eea
Here we have introduced a local magnetic field $h$ coupled to the 
impurity spin $S^z_f$, an on-site Coulomb repulsion $U$, 
and a hybridization $\Delta = \pi V^2 / 2$ to the 
conduction band orbital $c_{0 \sigma}$. 
Units are chosen as $\hbar = k_B = g = \mu_B = 1$.
Depending on the energy of the impurity level, $\epsilon_f$, 
different physical behaviour is realized. 
In the following, we focus on the symmetric 
($\epsilon_f = -U/2$) and mixed valence ($|\epsilon_f| \approx \Delta$) regimes. 
The conduction band (extending in the range $[-1,1]$) is already 
written in the \emph{linear chain} representation characteristic 
for NRG
\beq
H_0 = \sum_{n=0}^\infty \xi_n \left(c^\dagger_{n \sigma}\,
c^\pdag_{n+1\sigma} + h.c. \right)
\eeq
%
where the hopping matrix elements decay exponentially 
$\xi_n \sim \Lambda^{-n/2}$ due to a logarithmic discretization of the 
conduction band. 
This model -- while still a nontrivial many-body problem -- can now 
be solved by iterative diagonalization, keeping in each step 
only the lowest, most relevant levels. 
The number of iterations then corresponds to the temperature one 
is interested in according to 
$T_N = c \,\Lambda^{-(N-1)/2} $,  where $c$ is a constant 
of order one. For calculating static quantities, all necessary
information is thus obtained because only excitations on the scale 
$T_N$ are relevant. For dynamical properties, however, an additional energy
scale is provided by the frequency $\omega$ which may be much larger
than the temperature. Let us focus on the spin resolved single
particle spectral density 
\beq    \label{rho}
A_{\sigma}(\omega) = 
\sum_{n m} |\bra{m} f^\dagger_{\sigma} \ket{n}|^2 \, 
\delta\left(\omega - E_{m} + E_{n}\right)\, 
\frac{e^{-\beta E_m} + e^{-\beta E_n}}{Z}
\eeq
in the Lehmann representation where the $\ket{n}$ are the many-particle
eigenstates of $H$ and $Z$ is the partition function.
Obviously, spectral information at frequencies $\omega \gg T_N$ 
requires matrix elements between low-lying states and excitations
which in iteration $N$ are not available anymore (they have already been lost by truncation). 
To circumvent this difficulty, the following \mbox{``ad hoc''} procedure has been 
used so far: In calculating $A(\omega)$, expression (\ref{rho}) was 
simply evaluated in iteration step $N' \ll N$ where $T_{N'} \approx
\omega$, assuming that for an analysis of this spectral regime the low energy levels 
were described well enough. 
There is no rigorous argument to justify this assumption, as, for
example, the crossover to the strong coupling fixed point and the 
corresponding change in the ground state may occur at 
a much lower temperature scale $T_K \ll T_{N'}$. 
\begin{figure}[t]
\begin{center}
\epsfig{file=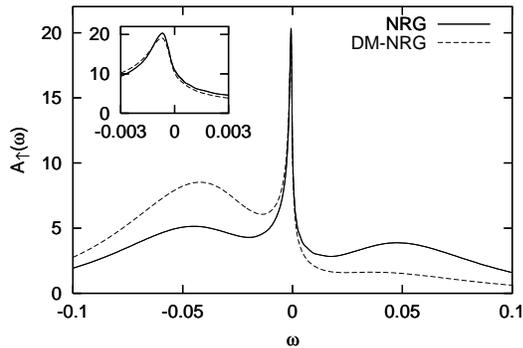,width=0.8\linewidth}
\end{center}
\caption{\label{fig:vergleich} Comparison of single particle spectral functions  
for the symmetric model ($\Delta=0.01$, $U=0.1$, $\epsilon_f = -0.05$) 
obtained by the method previously used (``NRG'') and the generalized procedure presented 
here (``DM-NRG'').
A small magnetic field $h=0.001$ has been applied to the impurity.
} 
\end{figure}
In fig.~\ref{fig:vergleich} we present results for the symmetric 
model (\ref{H_impurity}) at
$T=0$ which have been calculated in this way.
Without external field, one obtains the well-known three-peak
structure characteristic for a small Kondo temperature $T_K$. 
Switching on a small magnetic field $h = {\cal O}(T_K)$ only 
affects the quasiparticle peak, while the high energy spectrum 
is almost unchanged. This result is 
easily understood: In the iterations where the atomic levels are
calculated, the NRG procedure does not yet ``know'' about the 
tiny perturbation that eventually breaks the spin symmetry of the 
ground state. One can, however, easily verify that this result is incorrect: 
Calculating the ground-state magnetization $m$ (a static quantity)
directly as a thermodynamic expectation value 
$\langle (n_{f \uparrow} - n_{f \downarrow}) \rangle$
and comparing with the value derived from the spectrum
\beq    \label{magnetization}
m = \int_{-\infty}^0 d\omega\, A_{\uparrow}(\omega) - 
\int_{-\infty}^{0} d\omega\, A_{\downarrow}(\omega)
\eeq
the results do not agree (see table in fig.~\ref{fig:table}). 
Physically, the strong polarization of the impurity due to a 
small magnetic perturbation should suppress the upper 
atomic level because no particle excitations are possible anymore.
This suppression is drastically underestimated by the method used so far 
(indeed, in the limit of vanishing Kondo temperature $T_K$ it will 
not be seen at all).

In order to capture this effect it is clearly necessary to obtain the 
correct ground state \emph{before} calculating the spectra. 
This is achieved by the following two-stage procedure: 

1) NRG iterations are performed down to the temperature $T_N$ of
interest, in particular we choose $T_N \ll T_K$ to calculate 
ground-state properties. 
In each iteration step, we keep the information on the transformation 
between one set of eigenstates and the next, i.e.~we save the
corresponding unitary matrix. After obtaining the relevant excitations 
at temperature $T_N$ one can define the density matrix 
\beq
\hat{\rho} = \sum_m e^{-E^N_m/T_N}\, \ket{m}_{N} \bra{m}
\eeq
which completely describes the physical state of the system. 
In particular, the equilibrium Green's function can be written as
\beq   \label{new_Green}
G_{\uparrow}(t) = i \theta(t) {\rm Tr} \left(\hat{\rho}\, 
\left\{f^\pdag_{\uparrow}(t), f^\dagger_{\uparrow}(0)\right\}\right)
\eeq
\begin{figure}[t]
\begin{center}
\epsfig{file=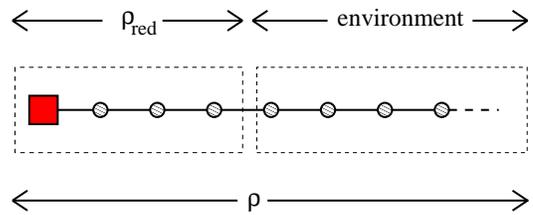,width=0.8\linewidth}
\end{center}
\caption{\label{fig:reduced_DM} Reduced density matrix obtained by
tracing out ``environment'' degrees of freedom of the chain.
} 
\end{figure}
2) Now we repeat the iterative diagonalization for the same
parameters. Each iteration step $N'$ yields the single-particle 
excitations (and matrix elements of $f^\dagger$) 
relevant at a frequency $\omega \sim T_{N'}$. 
But instead of using (\ref{rho}), we now 
employ (\ref{new_Green}) and evaluate the spectral
function with respect to the correct \emph{reduced density matrix}
\cite{Feynman 72}:
As depicted in fig.~\ref{fig:reduced_DM}, the complete chain is split 
into a smaller cluster of length $N'$ and an \emph{environment}
containing the remaining degrees of freedom. In the product basis 
of these two subsystems, the full density matrix has the form
\beq
\hat{\rho} = \sum_{m_1 m_2 \atop n_1 n_2} 
\rho_{m_1 n_1, m_2 n_2} \ket{m_1}_{\rm env} \ket{n_1}_{\rm sys} 
\bra{n_2} \bra{m_2}
\eeq
which is in general not diagonal anymore.
Performing a partial trace on the environment then yields the 
density submatrix 
\beq
\hat{\rho}^{\rm red} = \sum_{n_1 n_2} \rho^{\rm red}_{n_1 n_2} 
\ket{n_1}_{\rm sys} \bra{n_2}
\eeq
with 
\beq
\rho^{\rm red}_{n_1 n_2} = \sum_m \rho_{m n_1, m n_2}
\eeq
This projection is easily done using the previously stored 
unitary transformation matrices. 
Note that $\rho^{\rm red}$ -- defined only on the shorter chain -- 
contains all the relevant information about the 
quantum mechanical state of the \emph{full} system.
This concept has been applied very successfully in the 
density matrix renormalization group (DMRG) \cite{White 92}, where the 
projection on a smaller subsystem is essential for obtaining eigenstates 
of the model. In NRG, on the other hand, diagonalization 
can be performed directly due to the logarithmic discretization, 
but to describe the effect of the chain degrees of freedom on the impurity 
(or a small cluster) one has to determine $\rho^{\rm red}$.
In the following, we therefore refer to the calculational scheme
presented here as DM-NRG.

\begin{figure}[t]
\begin{center}
\epsfig{file=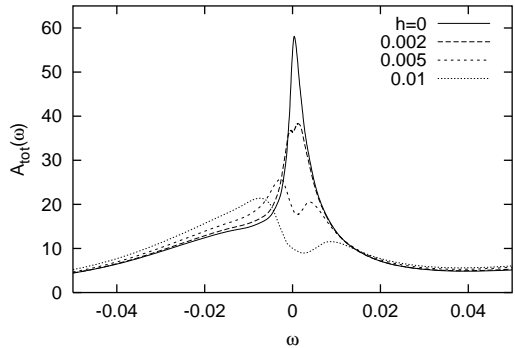,width=0.8\linewidth}
\end{center}
\caption{\label{fig:asymmetric_2} Total spectral density 
$A_{\rm tot}= A_{\uparrow} + A_{\downarrow}$ 
at zero temperature for
the asymmetric impurity with $\Delta=0.01$, $U=0.1$ and $\epsilon_f = -0.02$.
Note that upon increasing $h$ part of the spectral weight is shifted
to the upper atomic level (not shown). The total weight is constant 
with high accuracy.
} 
\end{figure}

In fig.~\ref{fig:vergleich} we compare the spectrum calculated by the
DM-NRG to the one obtained with the NRG version used so far in the literature
(the same number of levels has been used 
in both calculations). The strong shift of spectral weight due to the 
polarized impurity is now clearly seen, as well as a slight change 
in the height and shape of the quasiparticle peak. 
The magnetization has been calculated from (\ref{magnetization}) 
for different values of $h$ and is in good agreement 
with the static calculation (see fig.~\ref{fig:table}).
The remaining deviation of about three percent is due to an error 
in the total spectral weight.
\begin{figure}[t]
\begin{center}
\begin{tabular}{c@{\hspace{1cm}}c@{\hspace{1cm}}c@{\hspace{1cm}}c}
$h$ & $m_{\rm NRG}$ & $m_{\rm DM-NRG}$ & $m_{\rm direct}$ \\ \hline
0.0005 & 0.09  & 0.46  & 0.44 \\
0.001 & 0.16  & 0.61  & 0.60 \\
0.003 & 0.36  & 0.77  & 0.75 \\
0.005 & 0.49  & 0.81  & 0.80 \\
0.01  & 0.68  & 0.87  & 0.84 \\ \hline
\end{tabular}
\end{center}
\caption{\label{fig:table}Impurity magnetization obtained by 
different methods: from the spectrum ($m_{\rm NRG}$ vs. $m_{\rm DM-NRG}$) 
and as a thermodynamic expectation value ($m_{\rm direct}$).
Impurity parameters are chosen as $\Delta=0.01$ and $U=0.1$.
}
\end{figure}
\begin{figure}[t]
\begin{center}
\epsfig{file=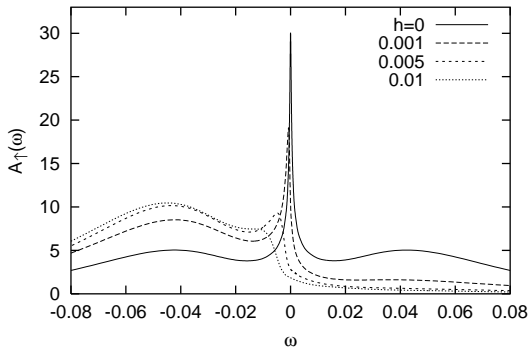,width=0.8\linewidth}
\end{center}
\caption{\label{fig:different_h} Shift of the spectral function with
increasing magnetic field at zero temperature. The impurity parameters are chosen as 
$\Delta = 0.01$ and $U=0.1$.} 
\end{figure}
The resulting field dependence of the spectrum is displayed in 
fig.~\ref{fig:different_h}. With increasing $h$, the Kondo resonance 
is suppressed and eventually merges with the lower atomic level. 
Regarding the total density of states (DOS)  
$A(\omega) = \sum_{\sigma} A_{\sigma}(\omega)$, the Kondo peak 
is split by the field and the DOS at the Fermi level
strongly reduced. This effect has been observed directly 
in measurements of the differential conductance through a quantum dot 
\cite{Goldhaber 98}.
\begin{figure}[t]
\begin{center}
\epsfig{file=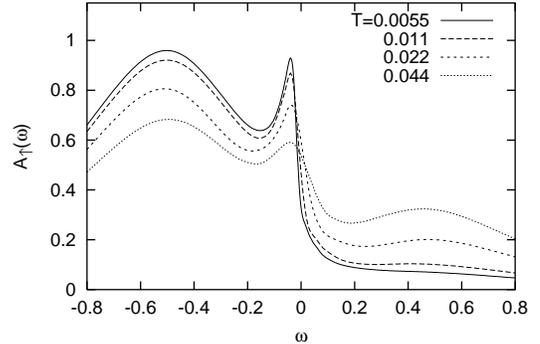,width=0.8\linewidth}
\end{center}
\caption{\label{fig:different_T} Temperature dependence of the
spectrum at $\Delta=0.1$, $U=1.0$ and $h=0.04$. Note that at 
frequencies far \emph{below} the temperature NRG does not yield any
information. In this region the curves are fitted.} 
\end{figure}
So far calculations have been at $T=0$. 
Upon increase of the temperature at a finite magnetic field, we expect a
reduction of the average impurity magnetization due to thermal fluctuations. 
As a consequence, particle excitations with polarization in the field direction
should gain spectral weight. 
In fig.~\ref{fig:different_T}, this effect is obvious: 
At temperatures $T \gtrsim h$, the asymmetry in $A_{\uparrow}(\omega)$ 
is strongly reduced. 
Note that in finite temperature NRG calculations, no spectral
information can be obtained at frequencies $\omega \ll T$. 
In this region data have to be fitted. 
This important fact will be discussed in detail in a subsequent publication. 
\begin{figure}[t]
\begin{center}
\epsfig{file=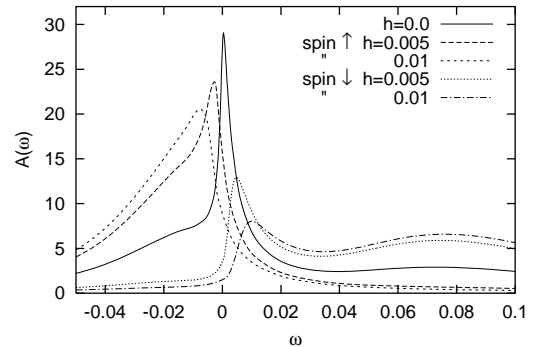,width=0.8\linewidth}
\end{center}
\caption{\label{fig:asymmetric_1} Spin dependent spectral density 
at zero temperature for
the asymmetric impurity with $\Delta=0.01$, $U=0.1$ and $\epsilon_f = -0.02$.} 
\end{figure}
Results for an asymmetric impurity close to the mixed valence regime are shown in
fig.~\ref{fig:asymmetric_1}. The almost complete 
shift of spectral weight to the particle (hole) sector 
is again observed for the two spin polarizations, which 
in this case are not symmetric anymore. 
In the total DOS (fig.~\ref{fig:asymmetric_2}), 
changes are less prominent. We merely observe a suppression of the 
quasiparticle peak and a redistribution of the corresponding 
weight to higher frequencies.

Comparing our findings to previous calculations, it should be pointed
out that up to now only the modified perturbation theory \cite{Takagi 99} 
and the Quantum Monte Carlo method (QMC) \cite{Sakai 99} have been
applied to calculate the impurity spectrum in a magnetic field.
The former is limited to small $U$,    
while QMC calculations have so far only been done in the 
mixed valence regime (and at temperatures $T \gtrsim T_K$)  
due to the increase in computing cost for the symmetric case. 
In a recent NRG calculation on the Kondo model \cite{Costi 00}, the problems discussed 
here did not occur due to the absence of atomic levels. 
Apart from these restrictions, we find qualitative
agreement with our results, which do not suffer from similar
limitations.  
In conclusion, we have presented a new method for calculating
dynamical properties at arbitrary temperatures within the numerical
renormalization group. It has been demonstrated that -- despite 
logarithmic discretization -- energy scale separation is in general 
not valid in the case of spectral quantities. This effect is neglected 
in the NRG scheme used so far in the literature. 
Within our generalized procedure (DM-NRG), based on the 
reduced density matrix, we can now account for changes 
in the ground state occuring at energies far below the external
frequency scale. 

The method introduced here has been applied to the Anderson impurity  
in an external magnetic field, which is of great interest 
in view of recent transport measurements of quantum dots. 
Nonperturbative $T=0$ studies have not been performed so far, mainly
because of technical difficulties in extending NRG to systems with broken spin symmetry.
Our spectral results are in excellent agreement with the sum rule
provided by the (static) magnetization. In the total density of states
we find the splitting and suppression of the quasiparticle peak 
which is also observed experimentally. 

Future applications of the DM-NRG include DMFT calculations for phases 
with long range order, where symmetry-breaking perturbations and 
their effect on the spectrum have to be treated reliably. 
In addition, more complex impurity systems including orbital degeneracy 
may be studied, which (due to the rapid advances in nanoscale 
preparation techniques \cite{Kergueris 99}) are of growing experimental interest.


The author would like to thank D.~Vollhardt, R.~Bulla and H.~Kontani 
for valuable discussions. This work was supported in part by 
the DFG through SFB 484.

\end{document}